# A dynamic intranet-based online-portal support for Computer Science teaching

K. Viswanathan Iyer
Dept. of Computer Science and Engg.
National Institute of Technology
Tiruchirapalli – 620015, India
email: kvi@nitt.edu

# Abstract

This paper addresses the issue of effective content-delivery of Computer Science subjects taking advantage of a university intranet. The proposal described herein for teaching a subject like Combinatorics and Graph Theory (CGT) is to supplement lectures with a moderated online forum against an associated intranet portal, which is referred to as a CGT-portal. The contents of a CGT-portal in a university intranet is required to be assembled by moderators and students during the progress of the CGT course. When completed at the end of a CGT course, a CGT-portal may be seen as a restricted view of the *Online Encyclopaedia of Integer Sequences* (OEIS: see http://oeis.org - the restriction can be with respect to sequences in OEIS that are directly relevant to say CGT). In the context of OEIS, an integer-sequence enthusiast experiences this cycle: *understand a page in OEIS-ponder over the contents-read afresh/refresh related content-suggest new additions to OEIS-wait for approval or rejection–* repeat this cycle. This experience can be imparted to students of a CGT course with the help of a CGT-portal. For organizing a CGT course, a first task is to partially create a miniature OEIS-like instructor-moderated CGT-portal in a university intranet. During the course of lectures and tutorials in CGT, students are asked to explore and contribute to the CGT-portal and these may be critically augmented or approved by instructors suitably, to find a place in the portal. Moderation also includes *feedbacks* (in many sense, a form of guidance) to students using/contributing to, the portal. By this, many concepts can be conveyed to the students in an interesting way with the desired results. It is pointed out that the dynamic nature of a CGT-portal promotes *active learning* philosophy the success of which depends on understanding the background and psychology of the student population. Some design guidelines associated with the building-up of a CGT-portal e.g., grouping of prerequisites of a logical page, knowledge-representation related observations, useful interfaces to the portal are also presented.

K.V.Iyer                                                                                                                                  NIT, Tiruchirapalli

# 1. Introduction

The paradigm shift from *education for the elites* to *education for the masses*, as evident from enrollments in institutions and from the growth in the number of educational institutes is of significant importance to the structure, purpose and social and economic roles of higher education in general. This has accelerated technological innovations in higher education in many places. The shift, from the viewpoint of what is referred to as non-traditional learners, is analyzed in Schuetze and Slowey(2002) with respect to the higher education scenario in ten developed nations. Therein, it has been stressed that in many developed countries policies on internet-based, distance and independent learning are issues for study in the context of e-learning. It implicitly follows that in a developed country or otherwise, the role of teachers and administrators and the attitudes of the student population are crucial to the success of any effort to build technology support to higher education. An important goal in technology support in teaching is to enrich the learning experience of students as is the case with intelligent tutoring systems. Conventional intranet-based education relies on high-end servers onto which lecture notes, handouts, hyperlinks to selected online-encyclopaedia pages, assignments and solutions etc. are preloaded by the instructors. To some extent these can cope-up with the rapid changes in a field like Computer Science and Engineering as a result of technological advances. However these approaches lack the feature of interactiveness as in computer-based problem-solving. Technological advances have prompted many teachers to promote *active learning* as a distinct and realizable paradigm to achieve stated learning outcomes. Success via active learning follows when among other things, the medium of communication and the interaction languages are well-established between the teacher-student group. Techniques in computer-aided-instruction-oriented systems are known to be teacher-assisted or otherwise. When newly introduced, often there is a *mismatch* between the ambitions of the teachers and educators versus the student adaptation of the activities offered. It has been pointed out that boredom and confusion precede undesirable behaviors among students although such computer-based learning systems are seen to be quite engaging in general and are useful when the number of students in a class is large. Thus a worthwhile research effort will be to reduce the said mismatch so as to minimize the undesirable effects of computer-based aids.

The suggestion described herein is a student-centric teacher-assisted use of a university intranet-web portal to be associated with a course that dynamically evolves with the progress of the course – this approach of describing and expanding knowledge, developing comprehension, analytic ability and inquisitiveness in a subject can be classified under computer-assisted learning in some sense. The approach outlined herein can be regarded as a positive step in enhancing a learning environment for undergraduate Computer Science teaching. For selective subjects (those involving intensive problem-solving tutorials and exercises), such as Discrete Mathematics or Combinatorics and Graph Theory (CGT) , the proposal is to supplement lectures with a teacher-moderated online forum (referred to as CGT-portal in the sequel) on a university intranet portal with restricted access to a CGT-class of students. In this the role of a teacher and his/her team is augmented by the role of students and it promotes collaborative learning as well, in computer-free classrooms. It also supports the future classroom models e.g., *split classrooms*.

To concretely understand the proposed approach, one can look at a restricted view of the Online Encyclopaedia of Integer Sequences (OEIS: refer to Sloane 2014) as a learning resource for CGT. (the restriction can be with respect to sequences in the OEIS that are directly relevant CGT). It may be noted that entries and modifications to the OEIS, contributed by integer sequences enthusiasts across the world are reviewed and then approved by a team of subject experts. A motivated student can explore the OEIS, fairly systematically and enhance his/her knowledge in Discrete Mathematics with additional teaching/learning support. An OEIS page is identified by a unique sequence number e.g., A158681 (see illustration 1). This sequence is on the Wiener indexes of the complete binary trees with k levels, the root being at level 0. As a prerequisite, this page requires notions such as *sequences, graphs, trees, distance in graphs, Wiener index, binary trees, recurrence relations.* A sequence enthusiast in the OEIS context will experience the following scenario: *understand a page entries in the OEIS - ponder over the contents - read afresh/refresh related content - suggest new additions to OEIS page/s - wait for approval or rejection –* repeat this cycle. During this process, a focused (e.g., with regard to Wiener indexes) and motivated sequence enthusiast will be able to acquire knowledge and will also be able to appreciate many aspects and alternative viewpoints on a specific topic. In a CGT course, for example, a student is expected to experience this process via an associated CGT-portal. The proposition of a CGT-portal for a CGT course, may be viewed as an eventual miniature form of OEIS although with many differences in content as well

**A158681  Wiener indexes of the complete binary trees with n levels, root being at level 0.**

```
4, 48, 368, 2304, 12864, 66816, 330496, 1579008, 7353344, 33583104,
151056384, 671219712, 2953068544, 12885491712, 55835820032, 240520790016,
1030797656064, 4398058045440, 18691721789440, 79164887531520,
334251639701504, 1407375101657088, 5910974963908608
```

| | |
|---|---|
| OFFSET | 1,1 |
| REFERENCES | R.Balakrishnan, K.Viswanathan Iyer, K.T.Raghavendra, "Wiener index of two special trees", MATCH Commun. Math. Comput. Chem., 57(2), 2007, 385-392. |
| LINKS | Table of n, a(n) for n=1..23. Index entries for linear recurrences with constant coefficients, signature (12,-52,96,-64). [R. J. Mathar, Sep 15 2010] |
| FORMULA | a(n) = (n+4)2^(n+1) + (n-2)2^(2n+2), n>0. G.f.: 4*x / ( (4*x-1)^2*(2*x-1)^2 ). [R. J. Mathar, Sep 15 2010] |
| EXAMPLE | For n=1, the complete binary tree with level 1 is P_{3} whose Wiener index is 4. |
| MATHEMATICA | LinearRecurrence[{12, -52, 96, -64}, {4, 48, 368, 2304}, 40] [Harvey P. Dale, Nov 05 2015] |
| KEYWORD | nonn,easy |
| AUTHOR | K.V.Iyer, Mar 24 2009 |
| EXTENSIONS | More terms from Harvey P. Dale, Nov 05 2015 |
| STATUS | approved |

Illustration 1:  A sample page in OEIS  (edited, hyperlinks removed)

as in the method of construction and usage, with one objective to provide *active learning experience* to the students in the CGT course with adequate traditional lectures in place.

The rest of the paper mentions briefly about the rationale behind the proposed portal-idea and the ways and means about how a similar portal can be developed in a university intranet for subjects related to Computer Science for undergraduate teaching. For understanding the illustrations, basic concepts in CGT is desirable (see for example, Bollobas, 1998; Chartrand and Zhang 2006; Andreescu and Feng, 2004).

## 2. Rationale for an intranet-portal for Computer Science teaching

In many ways, traditional learning is equated with systematic transfer and accumulation of knowledge. The connectivist learning-approach (Smith and Eng. 2013) as demonstrated by *MOOC*s appears to be a good technologically intensive alternative to address very large student population in a flexible way – in essence, the *process of learning* is given importance and knowledge creation is emphasized. Emphasis on the process of learning and acquiring knowledge is a feature in the proposed portal-design, a part of the course delivery. The following are the notable elements in a portal-design-based teaching:

- student participation in the portal including content-design.
- active references to the portal by instructors during regular lectures.
- feedbacks from portal moderators and instructors to student groups.

As mentioned in Iyer (2014), to organize a CGT course, one first task is to create partially, a CGT-potal in a university intranet environment, for example. The initial entries are due to an interest group with an identified moderating team. With the progress of course lectures and tutorials, students should be required to contribute to the portal and these may be augmented and approved by teachers and graduate students to find a place in the portal. An added advantage is that assignments and reading exercises can be against the portal. This is beneficial when the class strength is large, say 200 which is generally the case in many foundational courses. The moderators critically examine the student submissions and make suggestions – these are disseminated among the CGT-course students. Some of the activities such as supplying figures, formatting, suggesting computer codes, cross referencing pointers, verifications, pointers to other resources are assigned to a skilled groups of students. By this process, many concepts in a specific subject like CGT can be conveyed to the students in an interesting way with the desired results.

In the design of a CGT-portal, one emphasis is towards problem-solving competence. It may be recalled that in an unrestricted sense problem-solving competence is defined as the ability to solve cross-disciplinary and real-world problems by applying cognitive skills such as reasoning and logical thinking. Since this competence is regarded a desirable educational outcome, especially mathematics and science educators have focused on developing students' problem-solving and reasoning competence in their respective domain-specific contexts. Interaction of students and moderators against the CGT-portal background is expected to promote critical thinking among the target student population. One way to understand critical thinking (see Pascarella and Terenzini 1991, for example) is as given below.

It is a person's capability to:

(a) recognize core issues and assumptions in an argument e.g., definitions and properties associated with bipartite graphs
(b) make out important relationships e.g., understanding of alternative definitions of graph classes, structural properties of graphs

(c)  make correct inferences from data e.g., understanding lemmas, proofs
(d)  deduce conclusions from given information/data e.g., algorithmically test if a combinatorial object possesses certain specified features or characteristics
(e)  interpret whether conclusions are warranted on the basis of given data e.g., verify the correctness or incorrectness of a proof and provide a logic
(f)  evaluate evidence or authority e.g., scrutinize a conjecture, may be programmatically.

The comments and critisisms as well as suggested modifications that are offered by moderators of a CGT-portal are to be guided by these capabilities (a) through (f). It can be expected that student groups in the context of a CGT-portal content development will be implicitly exposed to critical thinking via arguments in CGT – this may be contrasted with traditional take-home assignments.

## 3. Evolution of a CGT portal for Computer Science teaching

Evolution of a CGT-portal and understanding the background and psychology of the student population in a CGT class and the importance of feedback to students are considered here. Some aspects are presented following Iyer (2014).

A CGT-portal can comprise several logical pages on a broad topic such as *Special Graphs and Combinatorial Objects* which is to be associated with a first-level course on CGT. A corpus of prerequisite keywords and phrases pooled into several perquisite boxes is to be associated with a CGT-portal. The corpus may be compared to a set of index terms at the back of a printed book. With each logical page of a CGT-portal one can associate primary attributes such as *definition(s), figures, constructions and algorithms to generate the objects, properties, importance in Combinatorics and Graph Theory, related objects or graphs, a more-to-explore set of references and/or reading materials, historical notes and significance, and remarks by students and instructors*. The properties and references will strive to capture a body of knowledge appropriate to CGT *per se* and to other courses requiring CGT as a prerequisite. Typical sources of knowledge include student submissions of assignment problems, copyright unprotected and free web/other resources, sources for which explicit permissions have been obtained, materials developed by instructors. As an active learning support, the dynamic nature – in terms of growth and restructuring - of the CGT-portal-idea is to be noted. By actively getting associated with the use and development of the CGT-portal, students in general will get involved to study, understand and assimilate more about the portal pages.

It is expected that enthusiastic teachers will always offer challenges via a CGT-portal. Teacher's challenges in the form of assignments and tests and open questions and students' responses are analysed and best understood against average grades representing different groups. Assuming that the possible student assessment grades are *A, B, C, D, E, F* it is possible to understand the level of knowledge and perseverance of the students as given below. Let  $t$  denote the term grade-point-average of a student so far in all subjects relevant to CGT, cleared by a student. We group the students based on $t$-value-mapped grade T and list a few descriptive attributes and conclusions for each group.

```
Group 1: T ε{A, B}
```
    Highly motivated in CGT, Able to solve a complete problem set by oneself, Able to pose well-formed problems, Self-initiated to use learning resources. Mostly applies self-regulatory procedures during learning.

Conclusion: Will be able to contribute entries to CGT-portal e.g., algorithms for generating or drawing graphs.

```
Group 2: T ε{C, D}
```
Fairly motivated in CGT, Able to understand the subject matter well with some effort, Completes exercises with rewards, Uses learning resources in a goal-oriented manner. Sometimes get the benefit of teacher-enforced regulatory mechanisms.

Conclusion: Will find CGT-portal as a learning resource and with guidance can contribute to page attributes.

```
Group 3: T ε{E, F}
```
Poorly motivated in CGT, Finds difficulty in getting started with the problems, Relies much on group activities, Needs significant effort to cross the subject. Requires teacher-enforced regulatory mechanisms.

Conclusion: Will be able to use the prerequisite boxes and will learn from CGT-portal at a slow pace.

We remark that scores obtained in an objective-type test in the topics e.g., prerequisites, related to CGT conducted by a CGT instructor at the beginning of a CGT course can augment a student's *t*-value.

The types of problems posed/arising through a typical CGT-portal page are as mentioned below.

a) Easily solvable by 85% of Group 1 students, 65% of Group 2 students.
b) Solvable with a knowledge of higher Mathematics developed through the CGT course and prerequisites.
c) Solvable via a programming language such as C, *Java* or *C++*.
d) Conjectures for which solutions or approaches are sought.
e) Problems and solutions from international contests such as Math Olympiad, ACM/ICPC.

To achieve the stated course outcomes, exercises in announced tests and in assignments in the context of a CGT-portal will test the following:

(i) Understanding of the definitions and key terms, properties etc. listed in a CGT-portal page - covering 60% Group 2 and all Group 3 students.

(ii) Understanding of notions and concepts via problems not immediately apparent from a CGT-portal page - covering 40% Group 2 and all Group 1 students.

For Group 2/3 students hints and suggestions in a tutorial session include
(a) Clarifications via explanations, maybe in a native tongue, explaining concepts/examples and/or via given assignments/quizzes.
(b) Supplementary reading in CGT and suggestions to explore specific CGT-portal pages and understand key-words and phrases possibly requiring additional reading of CGT-portal pages.

For Group 3 students counseling is also appropriate e.g., an advice to learn at a slower pace such as learning in a group under a supervision. Based on the effective usage of a CGT-portal and contributions to it, a teacher may expect that a Group 3/2 student will slide into Group 2/1 with the progress of a CGT course. It is apparent that a CGT-portal rightly implemented will promote problem-solving competence - understood as *the application of cognitive skills such as reasoning and logical thinking to solve cross-disciplinary and real-world problems* (refer Scherer and Beckmann 2014) – as a desirable educational outcome. It can be argued that the more-to-explore attribute and the prerequisites for a page will assist considerably in the teaching-learning process:

(a) Under the more-to-explore page attribute in an appropriate CGT-portal page can be a short write-up culled from contemporary literature, to provide possible use of Wiener-type topological indexes in Biochemistry for identification of lead drug candidates. This will add a further impetus for a Group 1

Computer Science student to discover more about total distance and other concepts such as cliques and independent sets in graph theory.

(b) It has been widely suggested (see Perrenet 2010, for example) that algorithm based problem-solving involves abstractions and it is an essential component in software development, a major concern for Computer Science students. An added activity of algorithm development and/or coding, for Computer Science majors, via a CGT-portal is therefore desirable. The more-to-explore page attribute in a CGT-portal will, among other things, point to pertinent combinatorial algorithms and graph algorithms at a high level – these may be to previously seen lectures and will provide more references for Group 1/2 students. In addition, problems suitably guided by the prerequisites are likely to suggest programming exercises e.g., computing the Wiener index of a graph given the weighted adjacency matrix suggesting say breadth-first search or Johnson's algorithm. In this process one may expect that Group 1 students will also propose new algorithms and program codes.

As seen in Perrenet (2010), the processes of *abstracting, concretizing, synthesizing* and *analyzing* are the mental faculties that are to be perhaps developed in teaching Computer Science – the extent to which a CGT-portal can contribute to these processes is an interesting question. In this context it can be noted that the Big Five personality factors (see, for example Ghorbani and Montazer 2015) are used to understand a learner's personality. Some pertinent questions against the Big Five factors in deploying and using a CGT-portal are as follows:

`Neuroticism:` As a personality dimension, neuroticism refers to stability and low anxiety at one end as opposed to instability and high anxiety at the other.
- Are Group 1 students' contributions to a CGT-portal the cause for neuroticism in Group 3 students?

`Extroversion-Introversion:` This is a characterization of a keen interest in other people and external events, and venturing forth with confidence into the unknown.
- Is the lack of probing and understanding the CGT-portal a source of introvertion among Group 3 students? Are the contents in the portal sufficiently accessible to Group 2 students?

`Openness:` This refers to how willing one is to make adjustments in notions and activities in accordance with new ideas or situations. Creativity, intelligence and imagination are associated with this factor.
- How open-minded are the Groups 2 and 3 students to the portal idea?

`Agreeableness:` This is associated with a flexible and cooperative attitude. It reflects how compatible people are with others and the ability to work in a group.
- Is agreeableness a problem among the students? For example, are the small number of Group 3 students getting isolated in the CGT class?

`Conscientiousness:` This refers to how much a person considers others when making decisions. It refers to traits of being thorough, in a task-oriented way, systematic.
- Are the Group 1 students paired with Group 2/3 students in upgrading the CGT-portal?

As pointed out in Ghorbani and Montazar(2015), efficiently identifying learners attributes is desirable in the development of e-learning environments and providing a personalized learning path.

In the CGT-portal context, explicit feedback to students is a necessary feature so that the uninitiated students can get a sense of guidance. Teacher feedbacks to students will be oriented towards encouraging student participation in the portal design and management. Feedback suggestions from instructors on students contributions to a CGT-portal is to be viewed with respect to the student categories and the overall goals of the program of which CGT is a part. Handling of student responses to teacher's comments is also an important aspect. The following are typical feedbacks to students:

`Feedback to Group-1/2 students:`
  i) Critical opinion
      e.g., *The teacher feels that you have not yet understood perfect graphs.*
      Context - *Student has made a contribution to perfect graphs.*
  ii) Suggestion to work more
      e.g., *Your submitted algorithm is correct but the result directly follows from the earlier discussions in the cited references.*
      Context – *Student has given an algorithm for finding terminal Wiener index of trees.*
  iii) Suggestion to read ahead
      e.g., *You are able to understand the pigeon-hole principle – you should read how the proof of 2-approximation algorithm for k-clustering, an NP-Hard problem, can use the principle.*
      Context - *Student has supplied her/his examples.*

`Feedback to Group-3/2 students:`
  i) Encouragement
      e.g., *The moderator has modified your suggested entry – check if the modification is correct so that your entry may be added to the portal.*
      Context – *Student has made a submission.*
  ii) Suggestion to refresh/update knowledge
      e.g., *You should checkout the given examples; first read pages xx and understand pre-requisite boxes pp.*
      Context - *Student has expressed his/her comments.*
  iii) Suggestion to go slow
      e.g., *Understand and illustrate the classic P.Erdös-G.Szekeres result via examples.*
      Context – *Students' queries on the lecture.*
  .

## 4. More design aspects of the CGT portal

A typical logical page in an A-CGT-portal will cover a *special graph* (e.g., graph of a hypercube) or it will describe a *graph-class* (e.g., perfect graphs) or a *specific concept* (e.g., permutations, combinations and binomial co-efficients). A page may thus describe binary trees, binomial trees, *AVL*-trees, spanning trees of complete graphs, *R-L-C* electrical networks, Flow networks, Reliability networks etc. A page may also be on say, graph representation; this will be covering *incidence matrix ,adjacency matrix, adjacency lists, sets of vertices and edges, diagrammatic representations, vertex-neighborhood pairs, sparse matrices and their representations, storage and redundancy in representations*. A typical skeletal page on r-regular graphs is outlined in more details in illustration 2. This page will have a set of associated keywords and phrases reachable from a tag say, `KEY-WORDS-PHRASES` – upon clicking, for example, this tag may open a prerequisite box to indicate the keywords, from a corpus, such as *undirected graphs, vertex degree, incidence matrix, diameter of a graph, Petersen graph* etc. If the r-regular graphs page is suggested as a selected reading assignment, a quiz question will be: give an example 4-regular graph that has an induced $K_4$.

As mentioned in Iyer (2014), the following two types of prerequisite boxes can be conceived:

**r-regular graphs**                                    **ID-RREGGRPHS**

**DEFINITION:**

An undirected graph is r-regular if for all vertex v in G we have *deg*(v) = r. Alternatively, for an undirected to be r-regular, the *max-degree* $\Delta(G) = r = \delta(G)$, the *min-degree*.

**A NEW DEFINITION:**

Consider the distinct positive integers $d_1, \ldots, d_r$. An undirected graph G is $(d_1, \ldots, d_r)$-regular if for all vertex v in G we have *deg*(v) in the set $D = \{d_1, \ldots, d_r\}$ and $|D|$, the minimum possible. *Note:* (r)-regular denotes r-regular.

**EXAMPLES:**

1. Graphs of a Set of edges, $C_k$, 3-cube, prism, Petersen graph, hypercube, odd graph are r-regular.

2. $K_{1,r}$ is (1,r)-regular; attach a pendant vertex to each vertex of a (r-1)-graph – this gives a (1,r)-regular graph.

**CONSTRUCTIONS:**

1. Let G be $(d_1, \ldots, d_r)$-regular with n vertices. Take two copies of G say $G_1$ and $G_2$. For each pair of corresponding vertices $v_1$ in $G_1$ and $v_2$ in $G_2$ introduce the edge $v_1 v_2$ between the two copies. This gives a new graph H with 2n vertices. H is $(d_2, \ldots, d_r)$-regular. One can iterate this step by taking a new G to be H. Eventually, this gives an r-regular graph.

2. Let G be r-regular with n vertices. Take two copies of G say $G_1$ and $G_2$. Let $u_1 v_1$ and $u_2 v_2$ be the corresponding edges (as in G) in these copies. Delete $u_1 v_1$ in $G_1$ and $u_2 v_2$ in $G_2$. Next add the edges $u_1 u_2$ and $v_1 v_2$ between $G_1$ and $G_2$. This gives a new r-regular graph H with 2n vertices. One can iterate this step by taking a new G to be H. This gives an infinite number of r-regular graphs.

**MORE-TO-EXPLORE:** Degree-Sum formula; Degree sequence of a graph.

**RELATED RESULTS:**

1. `Theorem [Hoffman(1963)]:`

   Let *K* be an *n* x *n* matrix where every element is 1. For a graph G with n vertices to be connected and regular a necessary and sufficient condition is that *K* is a linear combination of powers of *A(G)*, the adjacency matrix of G.

2. `Theorem [Hoffman and Singleton(1960)]:`

   Let G be an r-regular graph of diameter 2 (and therefore girth = 5), with n vertices where n = $r^2 + 1$. Then r= 2, 3, 7 or 57.

   `Note:` It is not known whether r=57 can be realized. The values r=2 (Pentagon),3 (Petersen graph),7 are realizable.

**REFERENCES:**

1. B. Bollobas, *Modern graph theory*, Springer, 1998.
2. G. Chartrand and P. Zhang, *Introduction to graph theory*, McGraw-Hill, 2006.

Illustration 2: A skeletal logical page in a CGT-portal

> `Type P1`: These comprise those keywords/phrases that are acquired in one or two sittings by a student. The definitions of these will be found in the box with a scrolling feature.
>
> `Type P2`: This encompasses keywords or phrases each of which point to 1-5 short pages of references e.g., planarity and embeddings, perfect graphs.

Although the precise distinction between type P1 and type P2 keywords is fuzzy, a rule of thumb is that type P1 keywords will be fully covered through the regular lectures. A color-encoding of a logical page's background or of the primary attributes will indicate which CGT-portal pages are part of the CGT course and which ones are considered outside the purview of the course. Alternatively and/or as an added feature a *degree of relevance* indicator computed as a weighted average based on the prerequisite boxes for each logical page of a CGT-portal is considered as a logical page attribute. The relevance will be to lecture notes and textbook chapters and sections.

Knowledge representation in a CGT-portal is another pertinent issue that will contribute to the ease of learning concepts. For more details on the observations made herein (Kolloffel et al. 2009) and other references therein should be consulted. Roughly, representational format mean diagrams, text, arithmetic and combinations of these. Diagrams are very useful in understanding complex material in a CGT-portal. This is understandable if we take examples of many proofs in Graph theory. However the precise nature of the advantages of a specific representational form in learning is attributed to complex interactions among the nature of the problem, resources used, student's ability, prior knowledge and practice time. It is generally assumed that features and characterictics of external representations will influence how one gets to details of information and how one can organize, interpret and remember presented information. An example is a two-colored graph presented as a picture vs. two lists of vertices one colored *red* and the other *black*; students will agree that correctness of the two coloring is more easily verified via the pictorial representation. As another example, presence of an odd cycle in small graphs can easily checked in a diagram than in a representation of two lists, one, the set of vertices and other, the set of edges. Examples such as these have suggested that although text (T) is equivalent to diagram (D) content-wise i.e., all the information in T can be inferred from D and *vice versa*, D is more effective in many situations as a learner can draw inferences quickly and easily from D. In literature this is referred to as *computational efficiency* of a representation as opposed to *informational efficiency* which says how representations organize information into data structures.

A few remarks on other external interfaces to a CGT-portal follow. A pedagogically useful interface to a CGT-portal is a graph drawing/manipulation program with the following major features:

(a) To draw a labeled graph on the plane given a specification of adjacency list/matrix.
(b) To mark all the vertices of a subgraph in grid on a plane, given the vertex set.
(c) To add all the edges of a graph given an edge-set and given a marking on a grid, the vertex-set.
(d) To redraw an already depicted graph in a grid on plane after specifying that a specified vertex needs to be positioned at a different grid point.
(e) To do the following graph operations (see also (d)):
- Add a new edge to a depicted graph by specifying two end vertices.
- Delete a vertex in a depicted graph along with its undirected incident edges.
- Add/mark in a grid a new vertex to an existing graph.
- Split an edge *uv* as the simple path *uxv* where *x* is new vertex.
- Graphically *contract* two vertices *u,v* along the edge *uv* in a depiction.

It easily follows that similar facilities can be provided in a graph drawing program for manipulating directed graphs. Other features such as presenting the vertices as circles and marking them with integers and/or filling them with prescribed colors may also be useful in understanding many concepts at the first-level. This can be seen through following examples:

Ex. 1. Seeing how $K_5$ or $K_{3,3}$ act as obstructors in non-planar input graphs.
Ex. 2. Designing algorithms for proper coloring or for finding min-cut wherein contraction is a useful operation.
Ex. 3. Radio-labeling of graphs, labeling graphs to find the bandwidth.

At the design level, the following are useful module-level interfaces, may be graphical user interfaces, to a CGT-portal:

i) A facility to search the portal contents via keywords and phrases to hit a set of page identifiers and prerequisite boxes.
ii) A facility within a CGT-portal to group a set of pages with a stated ordering of the selected pages so that a view may be presented to say Electrical Engineering majors or to Computer Science majors.
iii) A runtime-environment interface to execute stored codes may be in an interactive mode taking input from the keyboard or from files when required.
iv) A feedback section capturing history of contexts, instructor comments and student responses.
v) An electronic bulletin board interface to facilitate text-based communication between instructors and teachers with an optional smart-phone interface to the bulletin board.
vi) A traditional email interface to send/receive files as attachments.
vii) Links to locally stored free/copyright unprotected online electronic books and articles.
viii) An online manual on the description/usage of the CGT-portal and its facilities.
ix) A graph drawing program as mentioned above.

## 5. Concluding remarks

An outcome of maintaining students' records in a log, on the activities of contributing to page attributes or contributions as new page on the CGT-portal is that they can be suitably guided to credit further offered electives such as Coding Theory or Topics in Graph theory based on their major, and do guided project works in later semesters subject to their interests and career/program goals. At the end of a CGT course a developed CGT-portal can be edited further to serve as a profitable learning resource for various interest groups. The pedagogical value of a CGT-portal should be apparent but its implementation depends heavily on the preparedness of teachers as well. In a CGT course students' exploration of a CGT-portal will depend on attitudes, intellect and inquisitiveness. From the example illustrations it is possible to reason that a CGT-portal can judiciously contribute to the various levels of students' cognition as viewed from a psychological perspective, although a conclusion demands an experimental study and inference mechanisms.

From the viewpoint of freshers and some Group 2 and Group 3 students, didactical questions need further study. Many other aspects such as formats, presentation, ownership and management, copyright issues, incorporation of students' feedbacks and security issues in the context of portal development are not addressed here. It is instructive to investigate how the present proposal can be made to integrate well with some of the changing organization of higher education from traditional to lifelong learning models for teachers and students (see Schuetze and Slowey 2001): as opposed to restricted access, campus-cased and classroom-based, discipline-oriented and curriculum-centred imparting of basics in Computer Science the need is to support open access, off-campus/distance studies/self-learning, problem-solving & competence-oriented student-centred organization.

The implementation of a CGT-portal heavily depends on teacher-student collaboration, primarily driven by teachers. In general, the role of teachers in adopting information and communication technologies are known to be influenced by available time for professional development, support provided by schools, access to technology resources and teachers' feelings and beliefs about of technology. The research questions analysed in Kale and Goh (2014) are noteworthy in this context, although their work considers teaching with Web 2.0. Similar to Kale and Goh (2014) the open questions in accepting a CGT-portal-like idea are the following: (a) What are the teachers' attitude and experience with intranet usage? (b) What teaching style control do teachers have? (c) What are the factors that increase the appeal of a CGT-portal idea for teaching?

It may be argued that the portal-idea suggested is more towards specializing in one associated subject rather than breadthwise knowledge acquisition across the wide spectrum of subjects in Computer Science. This is to be weighed against further dependent courses in the program and the stated program goals and outcomes.